\documentclass{revtex4-2}
\usepackage{textgreek}
\usepackage{graphicx}
\usepackage{amssymb,amsmath}
\usepackage{graphicx}
\usepackage{color}
\usepackage{hyperref}
\usepackage{listings}
\usepackage{natbib}
\usepackage{upgreek}
\usepackage{setspace}

\newcommand{\s}{\text{S}}
\newcommand{\ii}{\text{I}}
\newcommand{\p}{\text{P}}
\newcommand{\h}{\text{H}}

\begin{document}
\title{Optical-parametric-oscillation-based $\chi^{(2)}$ frequency comb in a lithium niobate microresonator}

\author{N.~Amiune$^{1}$, D.N.~Puzyrev$^{2}$, V.V.~Pankratov$^{2}$, D.V.~Skryabin$^{2}$, K.~Buse$^{1,3}$, and I.~Breunig$^{1,3,*}$}
\affiliation{\footnotesize{
		\mbox{$^{1}$Laboratory for Optical Systems, Department of Microsystems Engineering - IMTEK, University of Freiburg, Georges-K\"ohler-Allee 102,}\\ 79110 Freiburg, Germany\\
		\mbox{$^{2}$Department of Physics, University of Bath, Bath BA2 7AY, England, United Kingdom}\\
		\mbox{$^{3}$Fraunhofer Institute for Physical Measurement Techniques IPM, Georges-K\"ohler-Allee 301, 79110 Freiburg, Germany}}
	\\$^*$optsys@ipm.fraunhofer.de}


\begin{abstract}
 Microresonator frequency combs based on the $\chi^{(3)}$ nonlinearity are nowadays well understood and making their way into different applications. Recently, microresonator frequency combs based on the $\chi^{(2)}$ nonlinearity are receiving increasing attention, as they promise certain benefits, but still require further study. Here, we demonstrate the generation of a $\chi^{(2)}$ frequency comb, initiated via optical parametric oscillation (OPO) in a lithium niobate mm-sized microresonator. By pumping at 532~nm with 300~\textmu W of power, we observe 1-THz-wide comb spectra around 1064~nm with degenerate and non-degenerate states. We also show that comb generation requires signal and idler waves to be degenerate in mode numbers and how the fulfillment of this condition can be identified from the temperature tuning curves. The results demonstrate the potential to directly generate frequency combs via OPO beyond 3~µm wavelengths in the mid-IR by puming in the near-IR region.
\end{abstract}
\maketitle
\section{Introduction}
Kerr frequency combs have significantly evolved during the last decade and have become a well understood system \cite{Kippenberg2018,Gaeta2019}. Nowadays, thousands of chip integrated soliton microcombs can be fabricated out of a single wafer\cite{Xiang21}, making this technology specially appealing for many applications such as  telecommunications, distance measurements (LIDAR), spectroscopy and quantum information processing among others. After the rapid growth of the activities dedicated to Kerr frequency combs, based on the $\chi^{(3)}$ nonlinearity, the possibility of microcomb generation using the $\chi^{(2)}$ nonlinearity has received more attention in the recent years, as it can provide new opportunities and advantages. Among them, there are the possibilities to use the linear electro-optic effect for stabilization and to work on more challeging wavelength regimes like the UV or the mid-IR\cite{Ricciardi2020}. The first realizations of $\chi^{(2)}$ frequency combs were done in bow-tie mirror cavities, started via second harmonic generation (SHG) and via optical parametric oscillation (OPO)\cite{Ulvila2013,Ulvila2014,Ricciardi2015,Mosca2016,Mosca2018}. It was shown that combs were formed around the pump and around the second harmonic frequency in the SHG scheme or around the sub harmonic frequency in the OPO scheme. The comb formation can be seen in a simple picture as a set of three-wave mixing cascaded processes between the pump and the second/sub harmonic frequency. Following these studies, $\chi^{(2)}$ combs were demonstrated via SHG in lithium niobate waveguides \cite{Ikuta2018} and bulk microresonators \cite{Szabados20,Hendry2020} with thresholds as low as 85~\textmu W\cite{Szabados20:2}. A soliton $\chi^{(2)}$ comb initiated via OPO was recently realized in an aluminium nitride chip-integrated microring resonator at telecom wavelengths, where $\chi^{(3)}$ nonlinear effects are also present\cite{Bruch2021}. On the theory side, there are several studies showing a variety of the soliton and non-soliton comb states in the microresonator parametric down-conversion~\cite{Villois2019,Podivlov2020,Smirnov2020,opexdvs} which still require a serious experimental investigation. In this work, we demonstrate for the first time the generation of a $\chi^{(2)}$ frequency comb initiated via OPO in a lithium niobate mm-sized microresonator. Moreover, we show that comb generation via OPO is not just given by operation at degeneracy, but also all interacting waves must correspond to the same transversal mode family, and we provide a way to verify the fulfillment of this condition via simple temperature tuning. The results allow us to define guidelines for future experiments, that may allow direct frequency comb generation in the mid-IR region by conveniently pumping at telecom wavelengths.

\section{General considerations for OPO-based $\chi^{(2)}$ comb generation}\label{sec:gen_pic}

\begin{figure}[h!]
\centering\includegraphics[width=0.8\textwidth]{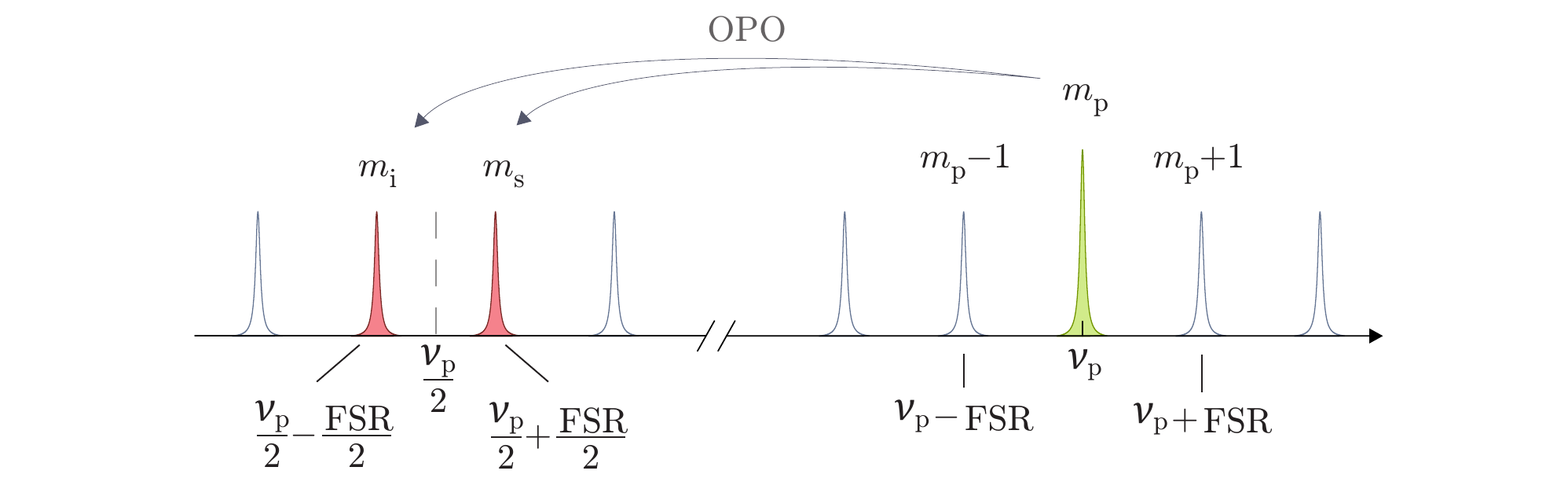}
\caption{ Optical parametric oscillation near degeneracy, where the signal and idler waves are spaced by one free spectral range. Internally pumped SHG is achieved only if $m_\text{s}=m_\text{p}/2 + 1/2$ and $m_\text{i}=m_\text{p}/2 - 1/2$}
\label{fig:cascades}
\end{figure}

The $\chi^{(2)}$ frequency combs are generated by a cascade of second order nonlinear processes, that take place back and forth between the pump and  subharmonic frequencies. These include optical parametric oscillation (OPO), second harmonic generation (SHG), sum frequency generation (SFG) and difference frequency generation (DFG)\cite{Ricciardi2020,pra}. The first down conversion step in the cascade picture can correspond to either the degenerate or near-degenerate OPO, which must fulfill energy and angular momentum conservation equations,
\begin{align}
\nu_\p&=\nu_\s+\nu_\ii,  \label{eq:ener_cons}\\
m_\p&= m_\s + m_\ii, \label{eq:mom_cons}
\end{align}
where $\nu_{\p,\s,\ii}$ are the frequencies and $m_{\p,\s,\ii}$ the azimuthal mode numbers for the pump ($\p$), signal ($\s$) and idler ($\ii$) waves. The situation of OPO with one free spectral range (FSR) spacing between signal and idler is sketched in figure \ref{fig:cascades}. There is, however, one important distinction of OPO operation near degeneracy, which depends on the transversal mode families involved for signal and idler.

The transverse modes of a whispering gallery resonator are determined by the mode numbers $p$ (zeros in polar direction) and $q$ (extrema in radial direction)\cite{Breunig16}.
If $q_{\p,\s,\ii}$ and $p_{\p,\s,\ii}$ are the transverse mode indices, then the signal and idler frequencies $\nu_{m_{\s}q_{\s}p_{\s}}$ and $\nu_{m_{\ii}q_{\ii}p_{\ii}}$, are fully defined by these three quantum numbers respectively.

\begin{figure}[h!]
\centering\includegraphics[width=0.8\textwidth]{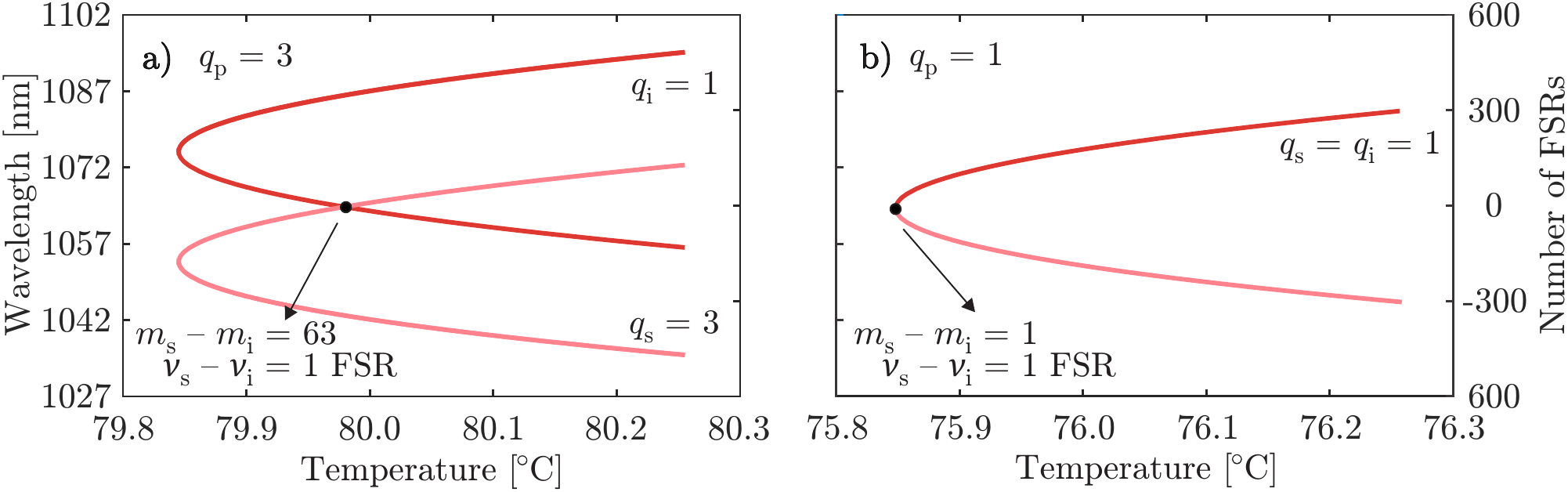}
\caption{Simulated temperature tuning curves for modes a) $q_\text{p,s,i}=(3,1,3)$, $p_\text{p,s,i}=(0,4,4)$ and for b) $q_\text{p,s,i}=(1,1,1)$, $p_\text{p,s,i}=(0,0,0)$ for a lithium niobate resonator with 1.25~mm radius.}
\label{fig:tuning_curves_simulation}
\end{figure}

The OPO temperature tuning curves near degeneracy can be calculated using equations \ref{eq:ener_cons} and \ref{eq:mom_cons} along with the single frequency condition for triply resonant OPOs\cite{sfc} and an approximation for the resonant frequencies including dispersion for the resonator geometry\cite{WGRmodes,Umemura2014}. 
As examples, the calculated tuning curve for mode numbers $q_\text{p,s,i}=(3,1,3)$, $p_\text{p,s,i}=(0,4,4)$ is shown in figure \ref{fig:tuning_curves_simulation}a), while the tuning curve for mode numbers $q_\text{p,s,i}=(1,1,1)$, $p_\text{p,s,i}=(0,0,0)$ is shown in \ref{fig:tuning_curves_simulation}b). We can then see that the tuning curves differ, depending on whether $q_\s$ and $q_\ii$ are equal or different.
More importantly, for near degenerate operation ($\nu_\s - \nu_\ii = 1~\text{FSR}$) we have for $q_\text{s}=q_\text{i}$ a difference between the azimuthal mode numbers of $m_\s - m_\ii = 1 $, while for $q_\text{s} \neq q_\text{i} $ a difference of $m_\s - m_\ii = 63$.

Let us now consider the next step in the cascade picture for comb generation, i.e. SHG of the signal and idler waves.
Assuming that the pump, signal and idler frequencies are close to resonance frequencies $\nu_{m,q,p}$, we can rewrite equations \ref{eq:ener_cons} and \ref{eq:mom_cons} separately for signal and idler as,
\begin{align}
\nu_{\s}&\approx \frac{\nu_\p+N \times \text{FSR}}{2},~\nu_{\ii}\approx \frac{\nu_\p-N \times \text{FSR}}{2}, \label{eq:freq}\\
m_\s&=\frac{m_\p+M}{2},~~~~~~~~~~m_\ii=\frac{m_\p-M}{2}\label{eq:mom},
\end{align}
with $N$ being the number of FSRs separating signal and idler frequencies, $N \times \text{FSR}=\nu_{\s} - \nu_{\ii}$ and $M=m_\s-m_\ii$.
If we consider SHG with these equations, we see on one hand that doubling of the signal and idler frequencies from Eq. \ref{eq:freq} would give rise to sidebands around the pump with a spacing of $N \times \text{FSR}$. On the other hand, doubling of the azimuthal mode numbers from eq. \ref{eq:mom} would provide phase-matching for sidebands $M$ modes away from the pump mode. From this analysis, it follows that SHG is possible only in the case that $M=N$, i.e. only when $q_\text{s}=q_\text{i}$.
It is therefore not sufficient to operate OPO near degeneracy for comb generation, but also the transversal mode numbers for signal and idler must be equal in order to achieve phase-matching for the cascaded processes that produce the comb lines.

\section{Experimental methods}
The manufactured whispering gallery resonator is made from a 300-\textmu m-thick 5\% MgO-doped congruent lithium niobate (CLN) wafer with the optic axis normal to the surface. From this wafer, we cut out a thin cylinder with 3~mm diameter using a femtosecond laser emitting at a wavelength of 388~nm with a 2~kHz repetition rate and 400~mW average output power. This cylinder is glued on top of a metal post for easier handling and further processing. Afterwards, we use the same laser to shape the cylinder and to obtain the desired resonator geometry. The resonator was manufactured with a major radius of $R = 1.25$~mm and a minor radius $r = 0.5$~mm. For the chosen geometry, the resonator's calculated free spectral range (FSR) is 16.7~GHz for 1064~nm ordinary polarization and 15.6~GHz for 532~nm extraordinary polarization. Finally, to obtain a good surface quality and minimize scattering losses, the resonator is polished manually with different diamond pastes with grain sizes down to 50~nm. The intrinsic quality factor was $1 \times 10^8$ at 1064~nm for o-polarization and $4 \times 10^7$ at 532~nm for e-polarization.

\begin{figure}[h!]
\centering\includegraphics[width=0.8\textwidth]{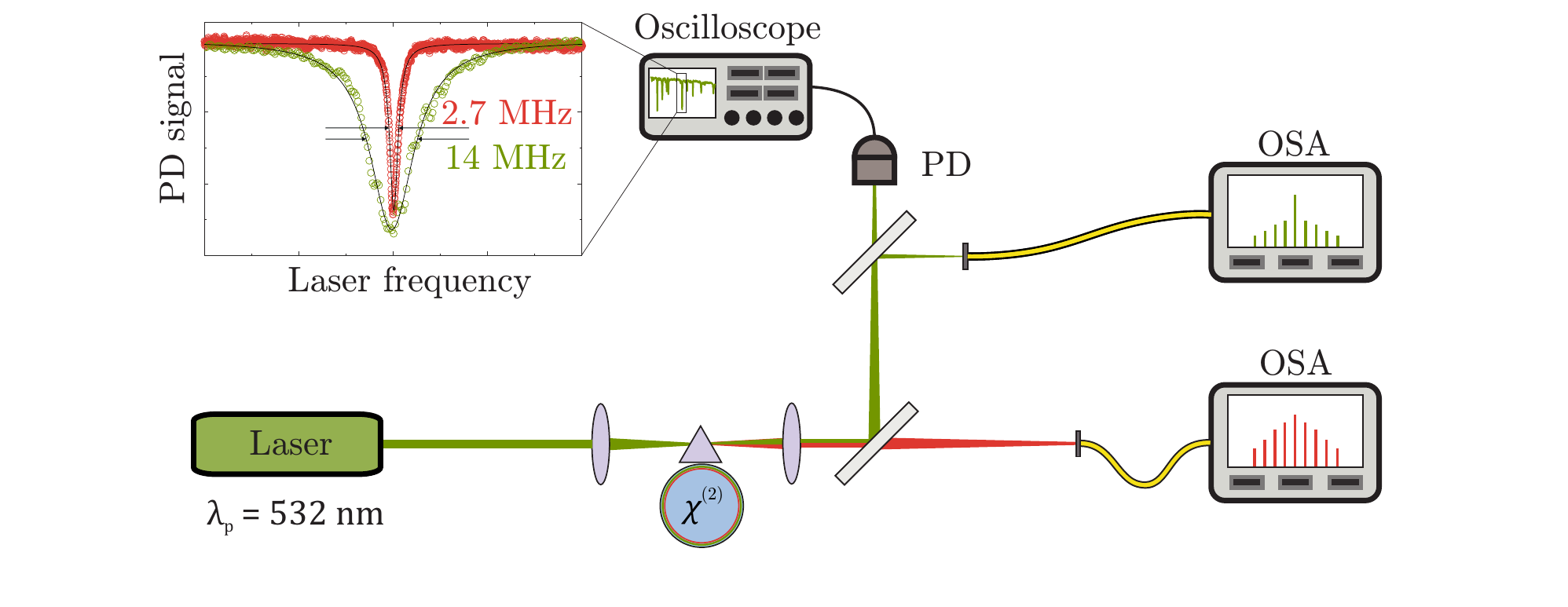}
\caption{Experimental setup and typical resonances at 1064 and 532~nm with their intrinsic linewidths. (OSA: Optical spectrum analyzer, PD: Photodetector)}
\label{fig:setup}
\end{figure}

The experimental setup is depicted in figure \ref{fig:setup}. An ALS 532~nm laser is used as a pump source. Light is focused and coupled into the resonator via evanescent coupling by using a diamond prism. We use birefringent phase matching with e-polarization for the pump beam and o-polarization for the signal and idler beams. The distance between the prism and the resonator can be adjusted by utilizing a piezoelectric translator, and the temperature of the resonator holder is stabilized with mK precision. Then, the outcoupled beam is split with a dielectric mirror separating the 532~nm and 1064~nm light. While the laser frequency is tuned by a few gigahertz, the outcoupled light beam at the pump frequency is monitored with a silicon photodetector to record the transmission and to identify different modes of the resonator. Subsequently, to carry out the experiment, we heat the resonator to temperatures around 75~°C, which is the calculated temperature for OPO degenerate operation in the fundamental mode. There, for different pump modes that provide OPO operation close to degeneracy, we measure the signal and idler output wavelengths, with a Blue-Wave grating spectrometer, as a function of temperature to retrieve the different tuning curves. During operation near the degeneracy, we use a Yokogawa AQ6370D optical spectrum analyzer to resolve the frequency comb spectra around 1064~nm and a Yokogawa AQ6373B optical spectrum analyzer to observe broadening of the pump at 532~nm. The spectra were recorded at incoupled pump powers of 60 and 300~µW.

\section{Results and discussion}
\subsection{OPO tuning curves and comb generation at low powers}
\begin{figure}[h!]
\centering\includegraphics[width=0.8\textwidth]{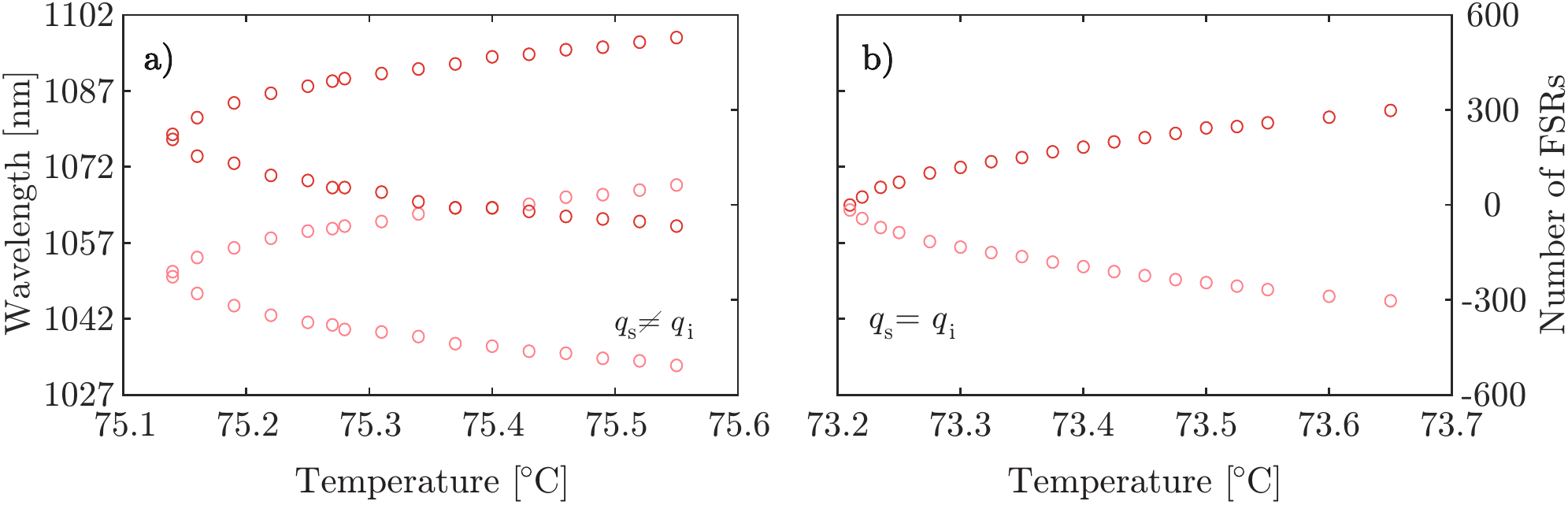}
\caption{OPO temperature tuning curves for two different pump modes in which signal and idler transverse mode numbers are different a) or equal b).}
\label{fig:tuning_curves}
\end{figure}

\begin{figure}[h!]
	\centering\includegraphics[width=0.8\textwidth]{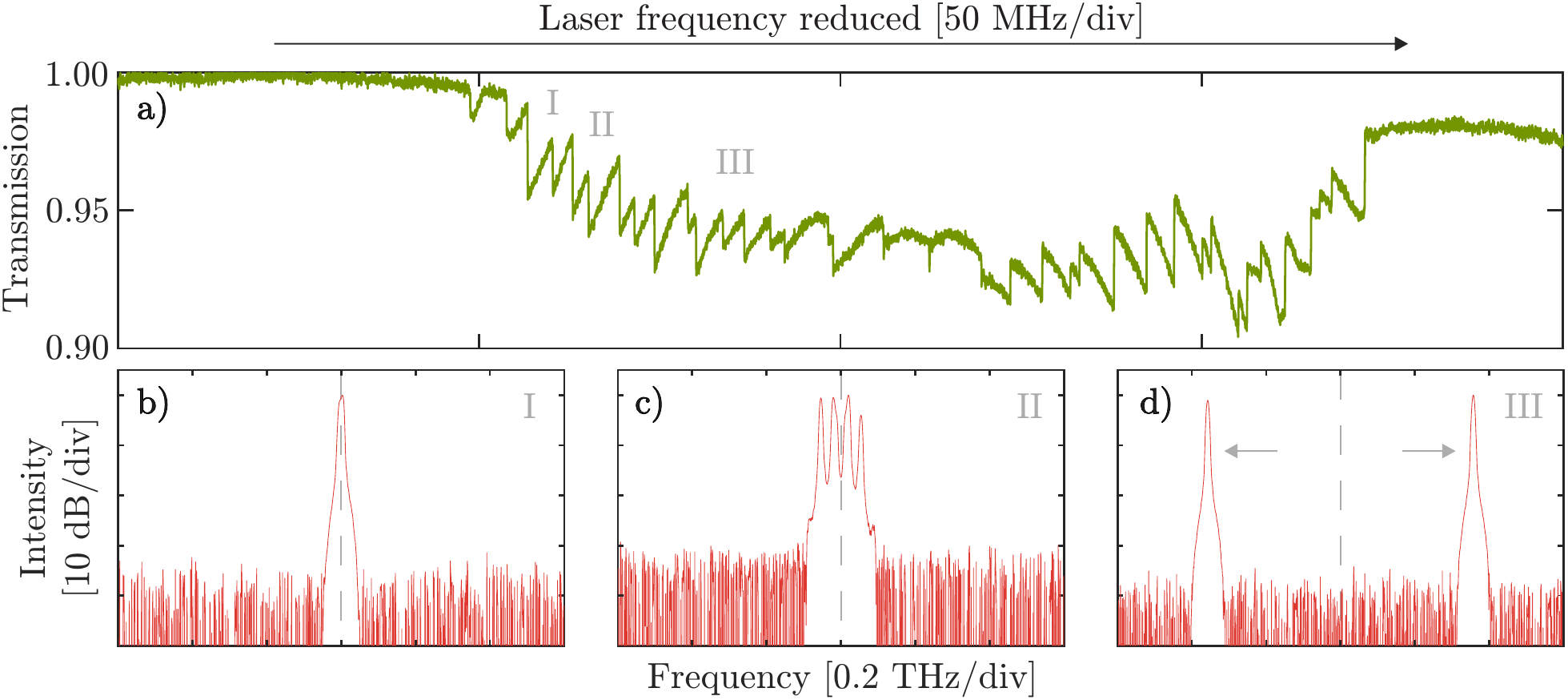}
	\caption{a) Scan of the pump laser frequency across the mode providing comb generation. b), c), d) OSA spectra around 1064~nm as the laser frequency was manually reduced across the resonace. The roman numbers are printed just to indicate in which order the spectra from figures \ref{fig:spectra_low}b) to d) are observed as the laser frequency was reduced manually, given that the amount of processes taking place prevents us from indicating this precisely. The gray dashed lines indicate the subharmonic frequency $\nu_\text{p}/2$.}
	\label{fig:spectra_low}
\end{figure}

We have measured the temperature tuning curves near degeneracy for two different pump modes, as shown in figure \ref{fig:tuning_curves}. We could identify two different types of tuning behaviors in good agreement with the calculations from section \ref{sec:gen_pic}: The tuning curve in figure \ref{fig:tuning_curves}a) corresponds to different transversal mode numbers for the signal and idler waves while the tuning curve in figure \ref{fig:tuning_curves}b) corresponds to equal mode numbers. Note that in case a), there is phase-matching for two different pairs of output signal and idler frequencies for a single temperature, each with a different threshold. When investigating operation at degeneracy in the case of different mode numbers, only stable OPO operation was observed with the OSA. This is expected, as this situation doesn't provide phase-matching for subsequent internally pumped cascaded processes, as discussed in section \ref{sec:gen_pic}. \\
However, when approaching degeneracy in the case of equal mode numbers, we observed the generation of extra sidebands even at powers as low as 60~µW. Figure \ref{fig:spectra_low}a) shows the transmission as the laser frequency was reduced across the resonance. Its shape differs from the typical triangular one of a thermally broadened resonance presumably due to the presence of several different OPO processes, mainly different spacings between signal and idler waves for the studied tuning curve. Note however, that also other OPO processes that provide output frequencies far from degeneracy are present. \\
As we tuned into resonance, we observed first OPO exactly at degeneracy, with signal and idler overlaping at $\nu_\text{p}/2$, as shown in figure \ref{fig:spectra_low}b). Then, by further reducing the laser frequency we entered a state with extra sidebands with 2 FSRs spacing between them as in figure \ref{fig:spectra_low}c). Afterwards, the system returned to the regular OPO state, as shown in figure \ref{fig:spectra_low}d), with signal and idler spacings getting larger, as the laser frequency was reduced, until the resonance was lost.\\ 

\subsection{Numerical modelling}\label{sec:sim_bath}
In the numerical simulations, we use the 
coupled-mode equations 
for the mode amplitudes $\psi_{\mu\h}$, $\psi_{\mu\p}$ as derived from the Maxwell equations in~\cite{josab} and implemented for the down-conversion set-up,
\begin{equation}\label{eq:equation:dynamics}
\begin{aligned}
i\partial_t\psi_{\mu\h} &= \left(
\frac{\delta+\varepsilon}{2}+D_{1\h}\mu+\frac{1}{2}{D_{2\h}\mu}^2\right)\psi_{\mu\h}
-\frac{i\kappa_\h}{2}\psi_{\mu\h}
-\gamma_2\sum_{\mu_1,\mu_2}{\widehat{\delta}}_{\mu,\mu_1-\mu}
\psi_{\mu_1\p}\psi_{\mu_2\h}^*~, \\
i\partial_t\psi_{\mu\p}&=
\left(\delta+D_{1\p}\mu+\frac{1}{2}D_{2\p}\mu^2\right)\psi_{\mu\p}- \frac{i\kappa_\p}{2}\left(\psi_{\mu\p}
-\widehat{\delta}_{0,\mu}\mathcal{H}\ \right) \\
&-\gamma_2\sum_{\mu_1,\mu_2}{\widehat{\delta} }_{\mu,\mu_1+\mu_2}\psi_{\mu_1\h}\psi_{\mu_2\h}\ ,
\end{aligned}
\end{equation}
where $t$ is time, and the subscripts 'P', 'H' correspond to the sidebands around the pump and half-harmonic fields, respectively. 
Here $\delta=\omega_{0\p}-\omega_\p$ is the detuning of  the pump photon from the cavity resonance 
at $\omega_{0\p}$, and $\varepsilon=2\omega_{0\h}-\omega_{0\p}$ is the frequency mismatch parameter.  $\omega_{0\p}$ is the resonance frequency with the even mode number, $m_\p$, and  $\omega_{0\h}$ is the frequency of the resonance with the number $m_\p/2$.  The repetition rate and dispersion parameters are set as  
$D_{1\p}/2\pi= 15.6$ GHz, 
$D_{1\h}/2\pi= 16.7$ GHz, 
$D_{2\p}/2\pi=-144$ kHz, and 
$D_{2\h}/2\pi= -65$ kHz.  
The parameters $\kappa_{\h}/2\pi=2.7$ MHz and 
$\kappa_{\p}/2\pi=14$ MHz are the intrinsic linewidths, and 
$\gamma_2/2\pi=300$MHz/$\sqrt{\text{W}}$ is the nonlinear coefficient proportional to the second-order susceptibility and inversely proportional to the mode area~\cite{josab}.
The Kronecker symbol is written as $\widehat\delta_{\mu,\mu'}$.
Assuming critical coupling, the pump parameter is given by
$\mathcal{H}^2=\mathcal{F_\p}\mathcal{W}/2\pi$, where
$\mathcal{W}$ is the pump laser power and $\mathcal{F_\p}$ is the finesse around $\omega_\p$. Here, $\mathcal{H}^2$, $|\psi_{\mu\h}|^2$, and $|\psi_{\mu\p}|^2$ have units of Watts and 
$\mu=0,\pm 1,\pm 2,\dots$ is the relative mode number.
In the model, we assume the same transverse mode structure  
within the 'H', and 'P' sideband groups. 

\begin{figure}[h!]
\centering\includegraphics[width=0.8\textwidth]{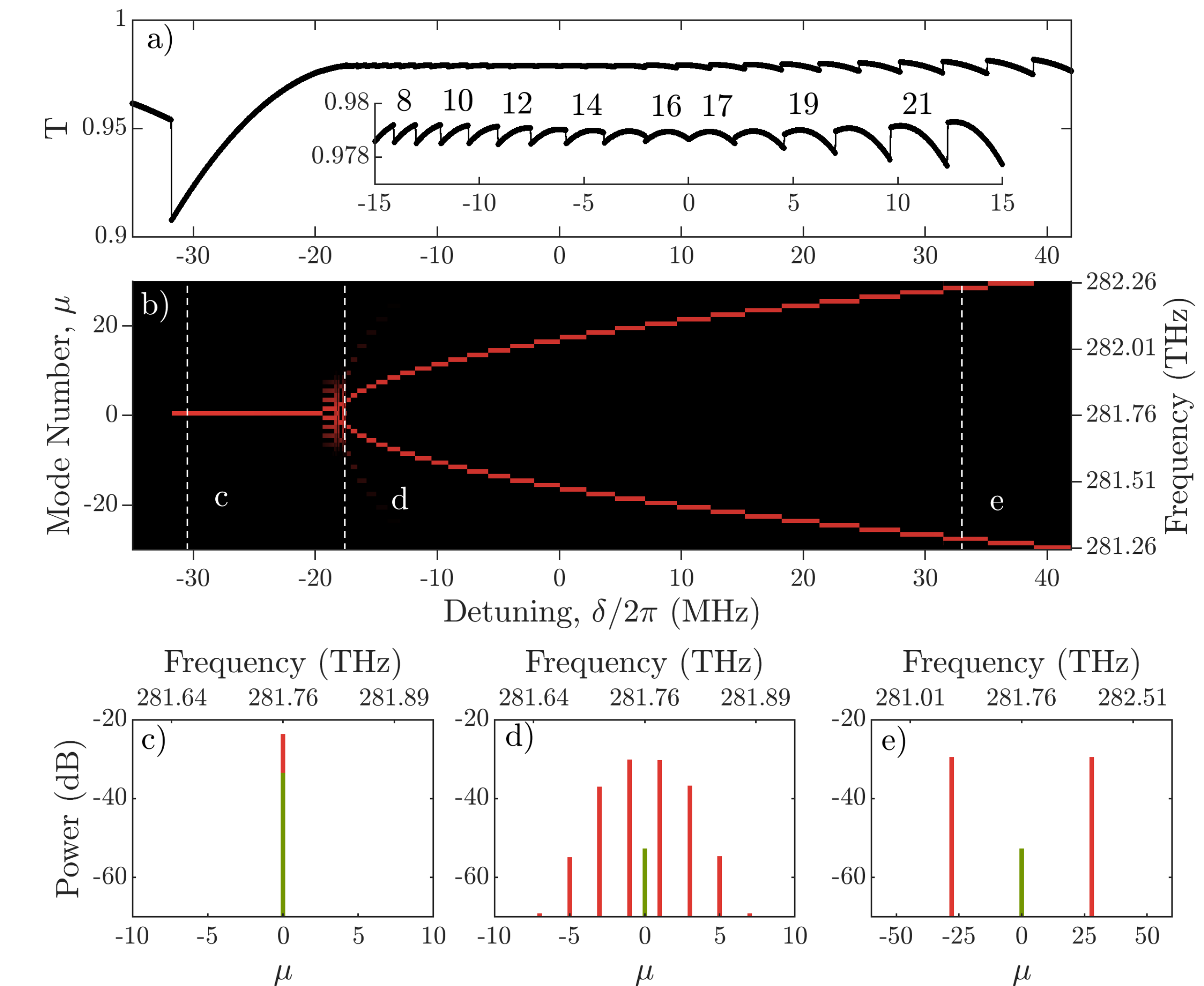}
\caption{(a) Numerically  modelled  resonator transmittance $T$. The inset  zooms on 
the detuning interval corresponding to the non-degenerate OPO states.  Panel (b) shows the spectrum of the half-harmonic field during the scan of the laser frequency $\omega_\p$.
The black background  corresponds to $-70$~dB. Panels (c-e)~show three examples of the spectra from~(b). The red bars correspond to the half-harmonic sidebands and the green one is the pump. The frequency labelling is for the half-harmonic field.  }
\label{fb}
\end{figure}

Our simulations are initialized by the no-OPO state, $\psi_\h=0$. The  resulting transmittance of the cavity, and the spectra generated when the pump frequency is scanned across the resonance, are shown in Fig.~\ref{fb}. The transmittance is calculated as the power dissipated over the roundtrip,
\begin{equation}
T= 1-\frac{2\pi}{\mathcal{W}}
\sum_{\mu}\left(
\frac{\left|\psi_{\mu\p}\right|^2}{\mathcal{F}_\p}+
\frac{\left|\psi_{\mu\h}\right|^2}{\mathcal{F}_\h}\right).
\end{equation}
The data shown are for the laser power $\mathcal{W}=60$~\textmu W and
$\varepsilon/2\pi= 17.7$~MHz. 

For sufficiently large negative detuning, the resonator prefers the degenerate OPO state, and the transmittance increases  with $\delta$ increasing, see
Fig.~\ref{fb}(c). As $\delta$ is approaching $-\varepsilon$, i.e., the half-harmonic detuning comes to resonance, a relatively narrow frequency comb is generated, see Fig.~\ref{fb}(d). The transition from the degenerate OPO to the comb state is similar to the one seen in the experimental data in Figs.~5(b), (c). Increasing $\delta$ further, $\delta>-\varepsilon$, 
brings the resonator to the non-degenerate OPO regime, where the generated side-band mode number $\pm\mu$ goes up with $\delta$ continuously, and can be well estimated from the condition 
\begin{equation}
\mu^2\approx \frac{\delta+\varepsilon}{|D_{2\h}|}.
\end{equation}
The full  derivation leading to this result will be presented elsewhere~\cite{conf}.
The transition to the non-degenerate OPO  also qualitatively agrees with the one observed experimentally, cf., Fig.~\ref{fb} and Fig. 5.

\subsection{Comb generation at higher powers}
\begin{figure}[h!]
	\centering\includegraphics[width=0.8\textwidth]{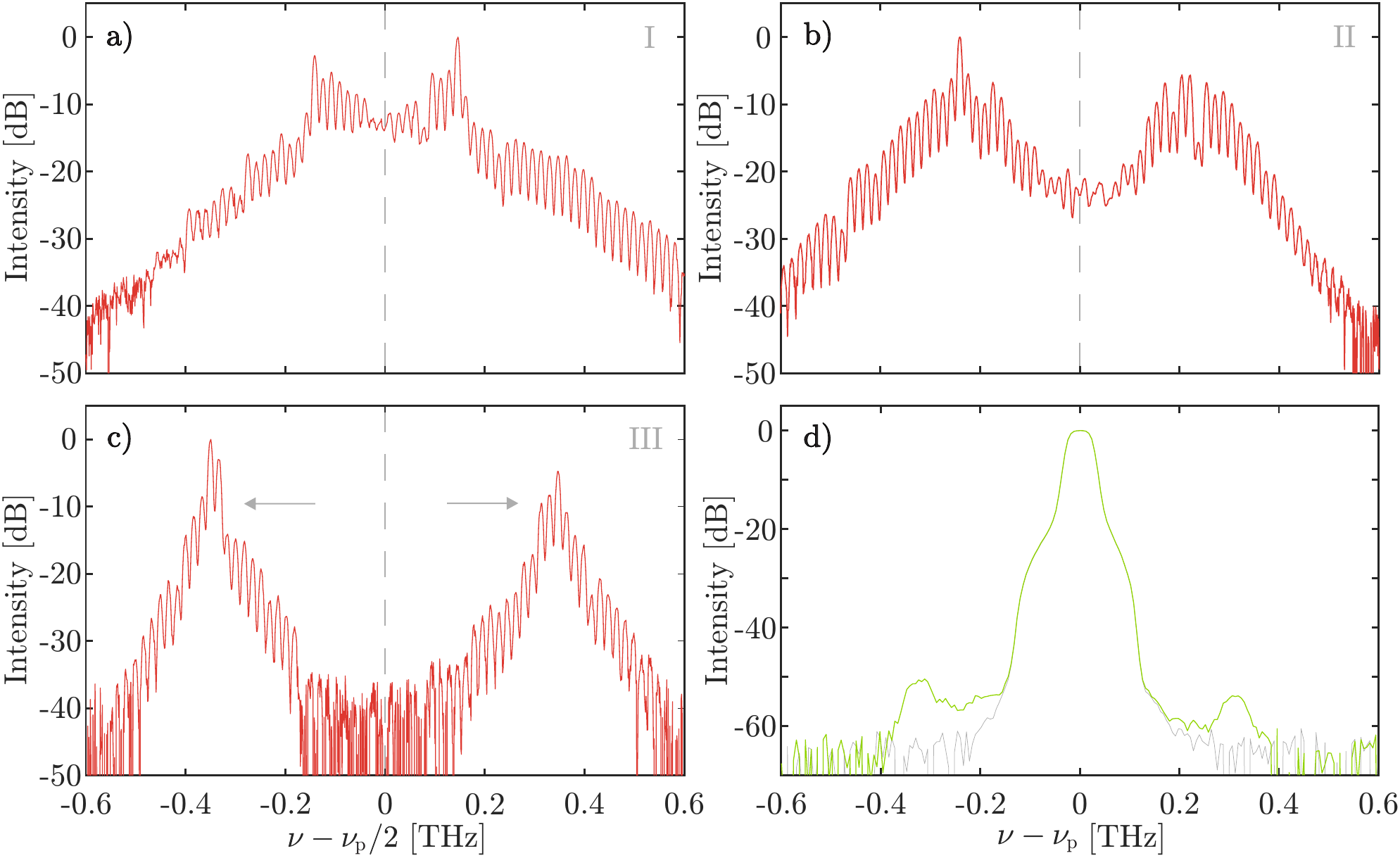}
	\caption{a), b), c): OSA spectra around 1064~nm as the laser frequency is tuned across the resonace; the gray dashed lines indicate the subharmonic frequency $\nu_\text{p}/2$. d): Spectra around 532~nm during state a); the gray continuous line indicates the pump laser when there's no OPO operation. }
	\label{fig:spectra_high}
\end{figure}

By increasing the incoupled pump power to 300~\textmu W we observe the generation of broader comb structures. These spectra are shown in figures \ref{fig:spectra_high}a)-c) for different detunings of the pump laser frequency, similarly as in figure \ref{fig:spectra_low}a). As the laser frequency was tuned into resonance, we observed first a degenerate comb state as in figure \ref{fig:spectra_high}a) with 1~THz width and lines spaced by 16.8~GHz corresponding to the FSR of the resonator at 1064~nm for o-polarization. By further reducing the laser frequency, the system transitioned to states with two non-degenerate combs as shown in figures \ref{fig:spectra_high}b)-c), similarly to previous observations \cite{Bruch2021}. Afterwards, only a pair of signal and idler peaks was detected, which again became more spaced until the resonance was lost. Generation of sidebands around the pump wavelength at 532~nm is shown in figure \ref{fig:spectra_high}d) as an example for the case of state a) in the NIR. Individual comb lines cannot be distinguished due to the limited resolution of the OSA in this wavelength range, but broadening of the pump can be observed. The low efficiency of the sidebands around the pump frequency is also seen in the simulations from section \ref{sec:sim_bath} which reveal that most of the power in the green is converted to the NIR.

Further investigation, going beyond our present scope, is needed to first find and then understand the high-power comb states in the numerical modelling. The parameter space of Eqs. (5) is vast and is far from being fully investigated by us. Also, the spectra measured can be sensitive to the parameters of the dynamical scan, and, on top of this, the nonlinear coupling between the different transverse mode families, which is disregarded here, is likely to play a more significant role as the pump power is increased. Beyond the model development, we should also note that the higher-order modes can be suppressed experimentally in our future work, by manufacturing a resonator with a geometry supporting only the fundamental mode. 

The results presented here can be considered as a first step towards OPO-based $\chi^{(2)}$ combs in lithium niobate, with the potential for future realizations integrated on a lithium-niobate-on-insulator chip where recently OPO was demonstrated\cite{Lu2021} and also high quality factors were achieved\cite{Gao2021}. Moreover, several microresonator studies argue that it is possible to generate two-color soliton combs via the half-harmonic generation~\cite{Ricciardi2020,Bruch2021,Villois2019,Smirnov2020,opexdvs,wlo}, if, for example, the FSR difference is brought down, which could be realized by pumping the resonator at around 675~nm with a quasi phase-matching structure. Moreover, this realization of a frequency comb via down-conversion offers new opportunities for the direct generation of combs in the mid-infrared region by conveniently pumping in the near-infrared.

\section{Conclusion}
In this work, we demonstrate for the first time the generation of a $\chi^{(2)}$ frequency comb initiated via optical parametric oscillation (OPO) in a lithium niobate mm-sized microresonator. We also show that frequency comb generation is achieved only in the case that signal and idler waves of the OPO are resonant for the same mode family, as well as an easy way to identify the fulfillment of this condition with the temperature tuning curves, providing guidelines for future experiments. This first demonstration of an OPO-based $\chi^{(2)}$ comb in a bulk microresonator opens up opportunities to be realized with lithium niobate chip-integrated platforms and to combine it with $\chi^{(3)}$ effects to get broader spectra. Moreover, this work also shows the potential to generate combs directly at longer wavelengths, being specially interesting the mid-IR region for applications, where OPO has already been shown with other materials such as CdSiP$_2$ or AgGaSe$_4$ pumped at telecom wavelengths\cite{Jia2018,Meisenheimer2017}.\\

\noindent {\bf Funding}
Horizon 2020 Framework Programme (812818, MICROCOMB);
UK EPSRC (2119373, DTP studentship).

\noindent {\bf Disclosures}
The authors declare no conflicts of interest. 

\noindent {\bf Data availability}
Data underlying the results presented in this paper are available from the corresponding author upon reasonable request.

\noindent {\bf Acknowledgements}
The authors thank H. Giessen (University of Stuttgart) and P. Del'Haye (MPI for the Science of Light) for support regarding experimental equipment.
\bigskip
\bibliography{LN_OPO_comb}

\end{document}